\documentstyle{article}
\begingroup\ifx\undefined\newsymbol \else\def\input#1 {\endgroup}\fi
\input amssym.def \relax
\input amssym
\newfont{\hr}{msbm10}
\newfont{\ams}{msam10}

\def \a {\alpha}

\def \d {\delta}

\def \det{{\rm det}}

\def \log {{\rm log}}

\def \ra {\rightarrow}

\def \be {begin {equation}}
\def \ee {end{equation}}
\def\tilde{\widetilde}
\def\bar{\overline}
\def\hat{\widehat}
\def\*{\star}
\def\({\left(}		
\def\){\right)}		
\def\[{\left[}		
\def\]{\right]}

\def\frac#1#2{{#1 \over #2}}

\def\d{\partial}

\def\2pi{\hbox{$2\pi i$}}

\def\dsl{\raise.15ex\hbox{/}\kern-.57em\partial}
\def\Dsl{\,\raise.15ex\hbox{/}\mkern-.13.5mu D}
\def\be{\beta}

\def\la{\lambda}	
\def\de{\delta}		\def\De{\Delta}
		
\def\sig{\sigma}	

		\def\CF{{\cal F}}

	\def\CN{{\cal N}}

\font\numbers=cmss12
\font\upright=cmu10 scaled\magstep1
\def\stroke{\vrule height8pt width0.4pt depth-0.1pt}
\def\topfleck{\vrule height8pt width0.5pt depth-5.9pt}
\def\botfleck{\vrule height2pt width0.5pt depth0.1pt}
\def\Zmath{\vcenter{\hbox{\numbers\rlap{\rlap{Z}\kern 0.8pt\topfleck}\kern
2.2pt
                   \rlap Z\kern 6pt\botfleck\kern 1pt}}}
\def\Qmath{\vcenter{\hbox{\upright\rlap{\rlap{Q}\kern
                   3.8pt\stroke}\phantom{Q}}}}
\def\Nmath{\vcenter{\hbox{\upright\rlap{I}\kern 1.7pt N}}}
\def\Cmath{\vcenter{\hbox{\upright\rlap{\rlap{C}\kern
                   3.8pt\stroke}\phantom{C}}}}
\def\Rmath{\vcenter{\hbox{\upright\rlap{I}\kern 1.7pt R}}}
\def\Z{\ifmmode\Zmath\else$\Zmath$\fi}
\def\Q{\ifmmode\Qmath\else$\Qmath$\fi}
\def\N{\ifmmode\Nmath\else$\Nmath$\fi}
\def\C{\ifmmode\Cmath\else$\Cmath$\fi}
\def\R{\ifmmode\Rmath\else$\Rmath$\fi}

\def\stackreb#1#2{\mathrel{\mathop{#2}\limits_{#1}}}
\def\Tr{{\rm Tr}}

\def\d{\partial}

\def\f{1\over}

\def\2{{1\over 2}}
\def\N2{${\cal N}=2$}

\def\be{ \begin{eqnarray} }
\def\ee{ \end{eqnarray} }

\def\d{\partial}

\def\bea{\begin{eqnarray}}
\def\eea{\end{eqnarray}}
\def\nn{\nonumber}

\def\beq{\begin{equation}}
\def\eeq{\end{equation}}
\def\ba{\beq\new\begin{array}{c}}
\def\ea{\end{array}\eeq}
\def\be{\ba}
\def\ee{\ea}
\def\stackreb#1#2{\mathrel{\mathop{#2}\limits_{#1}}}
\def\f{1\over}

\makeatletter
\newdimen\normalarrayskip              
\newdimen\minarrayskip                 
\normalarrayskip\baselineskip
\minarrayskip\jot
\newif\ifold             \oldtrue            \def\new{\oldfalse}
\def\arraymode{\ifold\relax\else\displaystyle\fi} 
\def\eqnumphantom{\phantom{(\theequation)}}     
\def\@arrayskip{\ifold\baselineskip\z@\lineskip\z@
     \else
     \baselineskip\minarrayskip\lineskip2\minarrayskip\fi}
\def\@arrayclassz{\ifcase \@lastchclass \@acolampacol \or
\@ampacol \or \or \or \@addamp \or
   \@acolampacol \or \@firstampfalse \@acol \fi
\edef\@preamble{\@preamble
  \ifcase \@chnum
     \hfil$\relax\arraymode\@sharp$\hfil
     \or $\relax\arraymode\@sharp$\hfil
     \or \hfil$\relax\arraymode\@sharp$\fi}}
\def\@array[#1]#2{\setbox\@arstrutbox=\hbox{\vrule
     height\arraystretch \ht\strutbox
     depth\arraystretch \dp\strutbox
     width\z@}\@mkpream{#2}\edef\@preamble{\halign
\noexpand\@halignto
\bgroup \tabskip\z@ \@arstrut \@preamble \tabskip\z@ \cr}%
\let\@startpbox\@@startpbox \let\@endpbox\@@endpbox
  \if #1t\vtop \else \if#1b\vbox \else \vcenter \fi\fi
  \bgroup \let\par\relax
  \let\@sharp##\let\protect\relax
  \@arrayskip\@preamble}
\def\eqnarray{\stepcounter{equation}%
              \let\@currentlabel=\theequation
              \global\@eqnswtrue
              \global\@eqcnt\z@
              \tabskip\@centering
              \let\\=\@eqncr
              $$%
 \halign to \displaywidth\bgroup
    \eqnumphantom\@eqnsel\hskip\@centering
    $\displaystyle \tabskip\z@ {##}$%
    \global\@eqcnt\@ne \hskip 2\arraycolsep
         $\displaystyle\arraymode{##}$\hfil
    \global\@eqcnt\tw@ \hskip 2\arraycolsep
         $\displaystyle\tabskip\z@{##}$\hfil
         \tabskip\@centering
    &{##}\tabskip\z@\cr}
\def\theequation{\thesection.\arabic{equation}}

\begin{document}
\begin{titlepage}
\setcounter{footnote}0
\begin{center}
\hfill ITEP/TH-57/97\\
\hfill FIAN/TD-14/97\\
\hfill hep-th/9710239\\
\vspace{0.4in}
{\LARGE\bf SUSY field theories, integrable systems}\\
\vspace{0.1in}
{\LARGE \bf and their stringy/brane origin -- II}
\\
\bigskip
\bigskip
\bigskip
{\Large A.Gorsky
\footnote{E-mail address: gorsky@vxitep.itep.ru}$^{,\dag}$,
S.Gukov
\footnote{E-mail address:
gukov@vxitep.itep.ru, gukov@pupgg.priceton.edu}$^{,\ddag,\dag}$,
A.Mironov
\footnote{E-mail address:
mironov@lpi.ac.ru, mironov@itep.ru}$^{,\S,\dag}$}
\\
\bigskip
\begin{flushleft}
$\phantom{gh}^{\dag}${\it ITEP, Bol.Cheremushkinskaya, 25,
Moscow, 117 259, Russia}\hfill\\
$\phantom{gh}^{\ddag}${\it Princeton University,
Princeton, NJ 08544, USA,\\
Landau Institute for Theoretical Physics,
Kosygina, 2, Moscow, 117940, Russia}\hfill\\
$\phantom{gh}^{\S}${\it Theory Department, P.N.Lebedev Physics
Institute, Leninsky prospect, 53, Moscow, 117924, Russia}\hfill
\end{flushleft}
\end{center}
\bigskip \bigskip

\begin{abstract}
Five and six dimensional SUSY gauge theories, with one or two
compactified directions, are discussed. The 5d theories with the matter
hypermultiplets in the fundamental representation are associated with the
twisted $XXZ$ spin chain, while the group product case with the
bi-fundamental matter corresponds to the higher rank spin chains.
The Riemann surfaces for $6d$ theories with fundamental matter and two
compact directions are proposed to correspond to the $XYZ$
spin chain based on the Sklyanin algebra.
We also discuss the obtained results within the brane and
geometrical engeneering frameworks and explain the relation to the toric
diagrams.
\end{abstract}

\end{titlepage}

\newpage
\section{Introduction}
\setcounter{footnote}0

In this paper we continue our discussion on Coulomb branch of gauge
theories in different dimensions, which
admit an exact treatment in the
low-enery sector. This is to say that we study the theories with eight real
supercharges (in accordance with the used terminology that means $N=2$ in four
dimensions and $N=1$ in $D=5,6$) with only four noncompact directions to have
non-trivial instantonic contribution and, therefore, a non-trivial
prepotential. This means that we consider the $5d$ theories with the
space-time $R^4 \times S^1$ and the $6d$ theories with the space-time
$R^4 \times T^2$. While
four-dimensional has been elaborated in \cite{ggm}, in this paper that is the
second one in the series, we give the detailed account of the results for the
higher (especially $5d$) dimensional theories, which have been announced in
\cite{ggm}. In particular, there have been stated the integrable structures
of the corresponding theories -- twisted $XXZ$ chain for the $5d$ theory and
$XYZ$ for the $6d$ one.

Let us note that since
the paper \cite{ggm} has been delivered, there appeared the paper
\cite{theisen} that dealt with $5d$ theories and, in particular, with the
corresponding spectral curves. The results of this paper \cite{theisen}
coincide with ours.

Thus, one of the main statements we are going to explain here is that the
$5d$ $SU(N_c)$ SUSY gauge theory with matter hypermultiplet in the fundamental
representation is described by the twisted $XXZ$ spin chain with $N_c$ sites.
In the second and the third sections of the paper we discuss this statement
in details. As a particular output, we obtain the spectral curves that
coincide with those of the paper \cite{theisen}. In the same section we also
briefly discuss the degeneration of the $XXZ$ spin chain to the relativistic
Toda chain, which has been already discussed in literature \cite{nikita,wdvv}
and corresponds to the pure gauge theory, i.e. bringing all the
hypermultiplet masses to infinity.

The spectral curves, obtained from the integrable system, as well as
generating 1-differential $dS$, allows one to define prepotential of the
theory within the framework of the Seiberg-Witten anzatz \cite{SW}.
Section 3 is devoted to the description of this procedure in the standard
$4d$ case and to its extension to the $5d$ case.

In fact, one of the important features of the $5d$ theories is that their
perturbative prepotential contains some
cubic (Chern-Simons) pieces \cite{S,IMS,Sei2,douglas,wdvv}. These cubic terms
can be generated by the 1-loop corrections \cite{IMS} as well as be included
into the bare lagrangian. It turns out that the integrability requires for
these terms to have the fixed coefficient and in the generic case saturates
the inequality found in \cite{IMS}. The same requirement can be equally
obtained from the associativity equations \cite{wdvv}. Since the cubic terms
are completely determined at the perturbative level, we specifically discuss
the perturbative prepotentials of the $5d$ theories in the third section.

After describing the integrable properties of the five-dimensional theories,
in very details,
in section 4 we come to the discussion of their stringy and brane origins.
To get eight real supercharges in $D=4,5,6$ we have to study
compactifications on the Calabi-Yau threefolds. The threefolds are
nearly the same (in the neighbourhood of the ADE-type singularity),
but the regions
of the Kahler moduli spaces are different. Thus, the four-dimensional
theory \cite{eng,V1} arise as we compactify type IIA theory
onto noncompact Calabi-Yau spaces and take the special limit of the large
base $P^1$'s and small $P^1$'s in the ALE fiber. Five-dimensional
theories (and all the possible twists) can be obtained via the
M-theory compactification. In this case, we are not
restricted to take this four-dimensional limit, and the resulting
theory is naturally defined in the $R^4 \times S^1$ space-time.
Remark that the spectral curves for the six-dimensional
theories are a local mirror to the del Pezzo singularity in
Calabi-Yau. In such a compactification near the singularity, the
effective strings in six dimensions become tensionless
\cite{SW6}. The
2-form, which they couple to in $d=6$ reduces to a 1-form in
five dimensions and to a 0-form in $d=4$. Under this dimensional
reduction, strings turn to point particles in $d=5$ and to
instantons in $d=4$. This explains how purely perturbative
dynamics in $d=5$ can lead to nonperturbative results in
the four-dimensional field theories \cite{nikita3}.

At last, after describing the $5d$ theories in very details, we sketch a
similar scheme for the $6d$ theories. In this case, we make some
more brief presentation, concentrating mostly on the essential points. In
particular, the main statement that we propose for this case is that the
integrable system corresponding to the $6d$ theories with the fundamental
matter is the $XYZ$ spin chain, with bare spectral torus being identified
with the space-time torus $T^2$.  We discuss the corresponding spectral
curves and the generating 1-differential $dS$ mainly following along the line
of the paper \cite{xyz}. The main difference of the $6d$ case as compared
with the $5d$ one is that we have not managed to find any degenerations of
the $XYZ$ chain to the (hypothetical) "elliptic" Toda chain (that would allow
one to consider the $SU(N_c)$ theory with the number of hypermultiplets less
than $2N_c$). This might match the essential feature of the $6d$
theories that they are self-consistant only if $2N_{c}=N_{f}$, otherwise, the
theory becomes anomalous \cite{6d1,6d2,6d3,6d4}.

When the present paper was almost complete, the related
paper by Aharony et al. \cite{ahk} have appeared. It has some overlaps with
our section 4. We give another evidence for
these proposal, in particular, explain the interplay among integrable
systems, brane configurations and geometric engineering.  To establish the
total equivalence in the last point with the results of Aharony et.al., one
has to interchange the axis in the toric polyhedra. Note that the grids are
indeed identical, while the rules for converting brane pictures into toric
varieties are different in both papers.

\section{Twisted $XXZ$ spin chain and $5d$ theories}
In this section we describe the integrable system behind the $5d$ theory,
that is, the inhomogeneous $XXZ$ spin chain, and obtain from this integrable
system the main ingredients necessary for the Seiberg-Witten (SW) approach:
the spectral curve of integrable system and the generating 1-differential
$dS$. Our presentation follows the line worked out in
\cite{gkmmm,xxx,xyz,ggm} and, therefore, we omit some details that can be
found in the cited papers.

\subsection{Twisted $XXZ$ spin chain}
The Lax matrix for the $XXZ$ ($SL(2)$) spin magnet
has the form
\be\label{l-gen}
L(\mu)\,=\,\left(
\begin{array}{cc}
\mu e^{S_0}-\mu^{-1}e^{-S_0} & 2S_-\\
2S_+ & \mu e^{-S_0}-\mu^{-1}e^{S_0}
\end{array}\right)
\ee
Thus, this Lax operator is given on cylinder.
In fact, from the point of view of integrable systems, some more natural
spectral parameter is $\xi$: $\mu=e^{\xi}$, and the Lax operator
(\ref{l-gen}) becomes clearly trigonometric.
This Lax operator is intertwined by the trigonometric $r$-matrix
\be\label{trigrmatrix}
r(\xi)={i\over\sinh\pi\xi}
\left(\sigma_1\otimes\sigma_1+\sigma_2\otimes\sigma_2+
\cosh\pi\xi\sigma_3\otimes\sigma_3\right)
\ee
so that the Poisson bracket of the Lax operators
\be\label{quadr-r}
\left\{L(\mu)\stackrel{\otimes}{,}L(\mu')\right\} =
\left[ r(\mu-\mu'),\ L(\mu)\otimes L(\mu')\right]
\ee
gives rise to the Poisson brackets of the algebra of $S_i$'s:
\be\label{pois}
\{S_+,S_0\}=\pm S_{\pm};\ \ \ \{S_+,S_-\}=\sinh 2S_0
\ee
The second Casimir function of this algebra is
\be\label{Casimir}
C_2=\cosh 2S_0+2S_+S_-
\ee

The non-linear commutation relations (\ref{pois}) are ones from
the quantum deformed algebra $U_q(SL(2))$.
This generalizes the fact that
$XXX$ magnet is described by the Poisson brackets that reproduce the
classical $SL(2)$ algebra (see details in \cite{ggm} and references therein).
In these Poisson brackets, the Plank constant $\hbar$ turns out to
be inessential and can be put equal to unity. In fact, it is proportional to
the radius of the space-time circle $R_5$ in the corresponding $5d$ SUSY
theory and can be easily restored by the replace of any generator
$S_i$ to $R_5S_i$. Hereafter, we omit $R_5$ from all the formulas excluding
special discussion in section 3.

Following the standard procedure now \cite{xxx}, one should consider
the chain with $N_c$ sites with Lax operators (\ref{l-gen}) associated with
each site and commuting with each other,
introduce the inhomogeneities $\xi_i$
that depends on the site of chain by the replace $\xi\to\xi+\xi_i$
and periodic boundary conditions.
The linear problem in the spin
chain, as usual reads as
\be\label{lproblem}
L_i(\mu)\Psi_i(\mu)=\Psi_{i+1}(\mu)
\ee
where $\Psi_i(\mu)$ is the two-component Baker-Akhiezer function.
The periodic boundary conditions are easily formulated in terms
of the Baker-Akhiezer function and read as
\be\label{pbc}
\Psi_{i+n}(\mu)=-w\Psi_{i}(\mu)
\ee
where $w$ is a free parameter (diagonal matrix).
One can introduce the transfer matrix shifting $i$ to $i+n$
\be\label{Tmat}
T(\mu)\equiv L_n(\mu)\ldots L_1(\mu)
\ee
which provides the spectral curve equation
\be\label{specurv}
\det (T(\mu)+w\cdot {\bf 1})=0
\ee
This equation generates a complete set of
integrals of motion.

The manifest form of the curve in the $XXZ$ case can be easily calculated
using manifest expression for the Lax operator (\ref{l-gen}):
\be\label{hrena}
w+{Q_{2N_c}\left(e^{2\xi}
\right)\over e^{2N_c(\xi-\xi_i)}w}=e^{-N_c(\xi-\xi_i)}P_{N_c}
\left(e^{2\xi}\right)
\ee
Note that by the change of variables $e^{2\xi}=\mu^2\equiv \lambda$, $w\to
e^{-N_c(\xi-\xi_i)}w$ this curve can
be transformed to the hyperelliptic form {\it in $\lambda$ variables}:
\be\label{sc2}
w+{Q_{2N_c}\left(\lambda\right)\over w}=P_{N_c}\left(\lambda\right),\
\ \ Y^2=P_{N_c}^2(\lambda)-4Q_{N_c}(\lambda),\ \ \ Y\equiv
w-{Q_{2N_c}\left(\lambda\right)\over w}
\ee
Certainly, in terms of the "true" spectral parameter
$\xi$ this curve looks considerably more tricky. However, one can use the
variable $\lambda$ taking into account that it lives {\it on the cylinder},
not on the sphere. In principle, it results in some subtleties that we
explain later (see the discussion of the residue formula).

In curve (\ref{sc2}), the polynomial
\be\label{212}
Q_{2N_c}(\lambda)=\prod_i \left(\lambda^2
-2C_{2,i}e^{-\xi_i}\lambda+e^{-2\xi_i}\right)\equiv
\prod_i\left(\lambda-e^{m_i^{(+)}}\right)
\left(\lambda-e^{m_i^{(-)}}\right)
\ee
defines the masses $m_i^{(\pm)}$ of the hypermultiplets. They turn out to be
functions of the second Casimirs $C_{2,i}$ of algebra (\ref{pois}) at $i$-th
site and the corresponding inhomogeneity. At the same time, the second
polynomial $P_{N_c})(\lambda)$ depends on gauge moduli.

Note that, by modulo this subtlety, the curve (\ref{sc2}) is similar to the
$4d$ curve. The difference, however, comes from the different generating
1-differentials $dS$ and turns out to be quite crucial.

\subsection{Degeneration to the relativistic Toda chain}
Now, in order to get a
natural degeneration to the pure gauge case that is described by the
relativistic Toda chain \cite{nikita} (see also \cite{wdvv}),
we should twist the $XXZ$ chain introduced above. The twisting
means just introducing some new parameters into the Lax operator, which are
the central elements of the Poisson bracket algebra. A special tuning of
these parameters allows one to match smoothly different regimes and limiting
cases of the $XXZ$ spin chain.

As it has been already indicated in our previous paper \cite{ggm}, there are
two equivalent ways to twist the integrable system. One of them, which was
applied in the $XXX$ case, is to multiply the Lax
operator by an arbitrary constant matrix $U$. This is possible in the $XXX$
case, since the corresponding (rational) $r$-matrix commutes with the tensor
product $U\otimes U$. The situation is much more restrictive in the
trigonometric case, when the $r$ matrix commutes only with the matrices $U$
of the very special form (say, it can be proportional to any one of the Pauli
matrices). However, in the trigonometric case we apply the second way of
doing \cite{kundu,Khar}, that is, we consider the Lax matrix of the general
form but with some prescribed dependence of the matrix elements on the
spectral parameter.  Then, the Poisson bracket (\ref{quadr-r}) dictates the
Poisson brackets of the matrix elements and, in particular, fixes some
coefficients to be centers of the Poisson bracket algebra.

More explicitly, we fix the Lax matrix to be of the form
\be\label{l-gener}
L(\mu)\,=\,\left(
\begin{array}{cc}
\mu A^{(+)}+\mu^{-1}A^{(-)} & B\\
C & \mu D^{(+)}+\mu^{-1}D^{(-)}
\end{array}\right)
\ee
and to be intertwined by the same $r$-matrix (\ref{trigrmatrix}). Then, up to
inessential total normalization, one can rewrite the Lax operator in the form
\be\label{l-general}
L(\mu)\,=\,\left(
\begin{array}{cc}
\mu e^{S_0}-\rho\mu^{-1}e^{-S_0} & 2S_-\\
2S_+ & \theta\mu e^{-S_0}-\nu\mu^{-1}e^{S_0}
\end{array}\right)
\ee
where $\rho$, $\nu$ and $\theta$ are
constants (centers of algebra).

In this case, the commutation relations are slightly changed to
\be\label{poisgen}
\{S_+,S_0\}=\pm S_{\pm};\ \ \ \{S_+,S_-\}={1\over 2}\left(
\nu e^{2S_0}-\rho \theta e^{-2S_0}\right)
\ee
and, respectively, the second Casimir function is equal to
\be
C_2={1\over 2}\left(\rho\theta e^{-2S_0}+\nu e^{2S_0}\right)+2S_+S_-
\ee
thus, the determinant of the Lax operator (\ref{l-general}) is still the
quadratic polynomial with the coefficient being the Casimir function.

Now we are ready to demonstrate how the limit to the relativistic Toda chain
can be done. To make the Lax operator (\ref{l-general}) looking more similar
to the ralativistic Toda Lax operator, we multiply it by the function
$e^{S_0}$.
In the whole chain it results into the factor
$e^{\sum_i^{N_c}S_{0,i}}$ that is the integral of
motion and can be put zero. In particular, in the relativistic Toda case this
integral is equal to the full momentum of the system. Although being so
inessential, this redefining of the Lax operator still requires to modify
some expressions \cite{Khar}. For instance, the new Lax operator
\be\label{laxnew}
L(\mu)\,=\,\left(
\begin{array}{cc}
\mu e^{2S_0}-\rho\mu^{-1} & 2S_-e^{S_0}\\
2S_+e^{S_0} & \theta\mu-\nu\mu^{-1}e^{2S_0}
\end{array}\right)
\ee
is intertwined by the new $r$-matrix (where $r$ is $r$-matrix
from (\ref{trigrmatrix}))
\be
r^{(tw)}=r+{1\over 2}\left(I\otimes\sigma_3-\sigma_3\otimes I\right)
\ee
that is called twisted. This name comes from the quantum
generalization of this matrix that can be obtained from the standard quantum
trigonometric $R$-matrix by twisting \cite{Khar}
\be
R^{(tw)}(u)\,=\,F_{12}(\theta)R(u)F^{-1}_{21}(\theta)'\ \ \
F_{12}\;\equiv\;F^{-1}_{21}\,=\,
\exp\Big\{\frac{1}{2}\big(\sigma_3\otimes 1-1\otimes \sigma_3\big)\Big\}
\ee
The important property of the Lax operator (\ref{laxnew}) is that its
determinant is no longer depending on only the Casimir function but also on
$e^{S_0}$. The product of this quantity over all the sites is certainly still
the integral of motion.

Now one can consider the particular reduction \cite{Khar}
\be
\theta=\nu=0,\ \ \rho=1
\ee
i.e.
\be
L(\mu)\,=\,\left(
\begin{array}{cc}
\mu e^{2S_0}-\mu^{-1} & 2S_-e^{S_0}\\
2S_+e^{S_0} & 0
\end{array}\right)
\ee
Taking into account the commutation relations (\ref{poisgen}), that is,
$\{S_+,S_-\}=0$, one can realize the algerba as $S_{\pm}=e^{\pm q}$, $S_0=q$,
$\{p,q\}=1$ and reproduce the standard Lax operator for the relativistic
Toda chain.

The spectral curve corresponding to this degenerated case can be easily
obtained now \cite{nikita,wdvv}
\be
w+{1\over w}={1\over\lambda^{N_c/2}}P_{N_c}(\lambda)
\ee

After constructing the degeneration procedure, one can always obtain from the
spectral curve (\ref{sc2}) describing the theory with $N_f=2N_c$ massive
hypermultiplets the curve with less number of massive multiplets
by degeneration of the general $XXZ$ chain at several sites (in complete
analogy with the $4d$ case, see \cite{ggm}). This results in the function
$Q(\lambda)$ in (\ref{sc2}) of the form
$\displaystyle{\prod_{\alpha}^{N_f}\left(\lambda-e^{2m_{\alpha}}\right)}$.

\subsection{$sl(p)$ spin chains}
The last issue we would like to discuss in this section is the higher $XXZ$
magnets that describe the $5d$ theory with the gauge group that is the
product of simple factors and matter hypermultiplets are taken in
bifundamental representations. In \cite{ggm} we discussed this connection in
the $4d$ theory, and now just briefly describe the corresponding system.

The Lax operator of this system that is $SL(p)$ $XXZ$ magnet is given by the
explicit formula
\be\label{laxhg}
L(\mu)=\sum_{i,j} e_{ij}\otimes L_{ij}
\ee
and
\be
L_{ii}=\mu e^{S_{0,i}}-\mu^{-1} e^{-S_{0,i}},\ \ \
L_{ij}=2S_{ji},\ (i\ne j)
\ee
Here $S_{0,i}$ are associated with the vectors $\epsilon_i$ realizing
simple roots as $\alpha_i=\epsilon_i-\epsilon_{i+1}$ and
$e_{ij}$ is the matrix with the only non-zero element $(i,j)$.
The generators $S_i$ satisfy non-linear Poisson brackets that can be
read off from the commutation relations of the quantum algebra $U_q(SL(p))$
like it has been done in (\ref{pois}). These Poisson brackets can be
certainly obtained from the quadratic relation (\ref{quadr-r}) with the
trigonometric $r$-matrix for the $SL(p)$ case \cite{FT}.

Now, using(\ref{laxhg}), one can obtain for the spectral curve the following
result
\be
w^p+{Q^{(1)}_{N_c}(\mu^2)\over\mu^{2N_c}}w^{p-1}+...+
{Q^{(k)}_{kN_c}(\mu^2)\over\mu^{2kN_c}}w^{p-k}+...+
{Q^{(p)}_{pN_c}(\mu^2)\over\mu^{2pN_c}}=0
\ee
Again, making change of variables, $\lambda\equiv\mu^2$,
$w\to \lambda^{-N_c}w$, one can finally obtain the
spectral curve
\be
w^p+Q^{(1)}_{N_c}(\lambda)w^{p-1}+...+
Q^{(k)}_{kN_c}(\lambda)w^{p-k}+...+
Q^{(p)}_{pN_c}(\lambda)=0
\ee
that again coincides with the $4d$ case.

Note that, specially degenerating this system, one can easily reproduce
the $(p,q)$-Web constructed in \cite{ahk} that corresponds to different toric
varieties. However, we prefer to consider here the general non-degenerated
system and return to this question later in section 4.

\section{Spectral curves and prepotentials: Seiberg-Witten anzatz}
In the previous section we described the Lax operators of the twisted
inhomogeneous $XXZ$ chain and constructed the corresponding
spectral curve. Now we describe how this data obtained from integrable theory
can be associated with the objects in SUSY theories, which reflects, indeed,
the content of the Seiberg-Witten anzatz.

\subsection{SW anzatz}
We start with the warm-up example of the $4d$ theory that has been originally
considered by N.Seiberg and E.Witten \cite{SW}.
For the \N2 SUSY gauge theory
the SW anzatz can be {\em formulated} in the following way.

First of all, one starts with a {\it bare} spectral curve,
with a holomorphic 1-form $d\omega $, which is elliptic curve (torus)
\be\label{torus}
E(\tau):\ \ \ \ y^2 = \prod_{a=1}^3 (x - e_a(\tau)), \ \ \
\sum_{a=1}^3 e_a(\tau) = 0,\ \ \ \ \ d\omega = \frac{dx}{y},
\ee
when the YM theory contains the adjoint matter hypermultiplet,
or its degeneration $\tau
\rightarrow i\infty$ -- the double-punctured sphere (``annulus''):
\be
x\rightarrow w\pm\frac{1}{w},\ \
y \rightarrow w\mp\frac{1}{w},\ \ \ \ \ d\omega = \frac{dw}{w}
\ee
otherwise. In particular, this latter possibility is the case
for the theory with the fundamental
matter hypermultiplets under discussion.

From the point of view of integrable
system, the Lax operator ${\cal L}(x,y)$ is defined as a function
(1-differential) on the {\it bare} spectral curve , while the
{\it full} spectral curve ${\cal C}$ is given by the Lax-eigenvalue equation:
$\det({\cal L}(x,y) - \lambda ) = 0$. As a result, ${\cal C}$ arises as a
ramified covering over the {\it bare} spectral curve:
\be
{\cal C}:\ \ \ {\cal P}(\lambda; x,y) = 0
\ee
In the case of the gauge group  $G=SU(N_c)$, the function ${\cal P}$ is a
polynomial of degree $N_c$ in $\lambda$.

The function ${\cal P}$ depends also on parameters (moduli)
$s_I$, parametrizing the moduli space ${\cal M}$. From the
point of view of integrable system, the Hamiltonians
(integrals of motion) are some specific co-ordinates on
the moduli space. From
the four-dimensional point of view, the co-ordinates  $s_I$ include $s_i$ --
(the Schur polynomials of) the adjoint-scalar expectation values $h_k =
\frac{1}{k}\langle\Tr \phi^k\rangle$ of the vector ${\cal N}=2$
supermultiplet, as well as $s_\iota = m_\iota$ -- the masses of the
hypermultiplets. One associates with the handle of ${\cal C}$ the gauge
moduli and with punctures -- massive hypermultiplets, masses being residues
in the punctures.

The generating 1-form $dS \cong \lambda d\omega$ is meromorphic on
${\cal C}$ (hereafter the equality modulo total derivatives
is denoted by ``$\cong$''). The prepotential is defined in terms of the
cohomological class of $dS$:
\be
a_I = \oint_{A_I} dS, \ \ \ \ \ \
\frac{\partial F}{\partial a_I} = \int_{B_I} dS \nn \\
A_I \circ B_J = \delta_{IJ}.
\label{defprep}
\ee
The cycles $A_I$ include the $A_i$'s wrapping around the handles
of ${\cal C}$ and $A_\iota$'s, going around the singularities
of $dS$.
The conjugate contours $B_I$ include the cycles $B_i$ and the
{\it non-closed} contours $B_\iota$, ending at the singularities
of $dS$ (see \cite{wdvv} for more details).
The integrals $\int_{B_\iota} dS$ are actually divergent, but
the coefficient of divergent part is equal to residue of $dS$
at particular singularity, i.e. to $a_\iota$. Thus, the divergent
contribution to the prepotential is quadratic in $a_\iota$, while
the prepotential is normally defined {\it modulo} quadratic combination
of its arguments (which just fixes the bare coupling constant). In
particular models $\oint_{A,B} dS$ for some conjugate pairs of contours are
identically zero on entire ${\cal M}$: such pairs are not included into our
set of indices $\{I\}$.

Note that the data the period matrix of ${\cal C}$ $T_{ij}(a_i)=
{\partial^2 F\over \partial a_i\partial a_j}$
as a function of the action variables $a_i$
gives the set of coupling constants in the effective theory.

The most important property of the differential $dS\cong\lambda{dw\over w}$
is that its derivatives w.r.t. moduli gives holomorphic differentials on
${\cal C}$ (see \cite{wdvv}).

The general scheme presented above can be almost literally transferred to the
$5d$ \N2 SUSY gauge models
with one compactified dimension. It can be described through
involving trigonometric $r$-matrices and $L$-operators instead
of rational ones, i.e. coming from Yangian to affine algebras. It means
that now it is natural to consider the {\it both} parameters $\lambda$ and $w$
lying on the cylinder.

Similar arguments imply that, instead of differential $dS^{(4)}=\lambda
d\log w$, one now has to consider the differential (see also
\cite{nikita,wdvv})
\be\label{dSRTC}
dS^{(5)} = \xi{dw\over w} \sim \log\lambda {dw\over w}
\ee
so that, despite the similarity of the $5d$ spectral curves with $4d$
ones, the periods of $dS^{(5)}$ are different from those of $dS^{(4)}$.
Note that the derivatives of this differential w.r.t. moduli again give
holomorphic differentials.

In fact, one can interpret the $5d$ theory as the $4d$ theory with infinite
number of (Kaluza-Klein) vector multiplets with masses $M_n=n/R_5$ and
infinite number of analogous (Kaluza-Klein) fundamental multiplets.
Then, one can equally consider the just described $5d$ picture, either the
$4d$ picture that involves Riemann surface presented by infinite order
covering (see the spectral curve equation in variables $\xi$ (\ref{hrena}))
with infinitely many punctures. This latter picture can be effectively
encapsulated in the usual hyperelliptic Riemann surface (\ref{sc2}) with
finite number of punctures, but, as a memory of infinitely many multiplets,
the spectral parameter $\lambda$ now lives on the cylinder. Meanwhile, the
differential $dS^{(5)}$ now evidently should be of the form (\ref{dSRTC})
that "remembers" of its $4d$ origin.

\subsection{Perturbative prepotential in $5d$ theory}
Now we discuss the perturbative part of the prepotential for
the $5d$ $SU(N_c)$ theory with $N_f\ge 2 N_c$ massive fundamental
hypermultiplets and compare this with the answers appeared in literature.

The explicit expression for the perturbative prepotentials can be derived
using formulas (\ref{defprep}) and (\ref{dSRTC}).
In the perturbative limit, the hyperelliptic curve (\ref{sc2}) turns into
the sphere
\be\label{pertcurve}
w={P_{N_c}(\lambda)\over\sqrt{Q_{N_c}(\lambda)}}\equiv
{\prod_i^{N_c}\left(\lambda-\lambda_i\right)\over\sqrt{
\prod_i^{N_f}\left(\lambda-\lambda_{\alpha}\right)}}
\ee
with $\lambda_i=e^{a_i}$, $\sum_i^{N_c}a_i=0$, $\lambda_{\alpha}=
e^{m_{\alpha}}$.
The perturbative differential $dS$ is of the form
\be\label{pertdS}
dS = \log\lambda\  d\log\left(\frac{P(\lambda)}{\sqrt{Q(\lambda)}}\right)
\ee

In the simplest case of the pure gauge $SU(N_c)$ Yang-Mills
theory, i.e. $Q(\lambda)=Q_0(\lambda)=1$, the perturbative part of the
prepotential is of the form \cite{nikita,wdvv}
\be\label{FARTC}
{\cal F}={1\over 4}\sum_{i,j}\left({1
\over 3}\left|a_{ij}^3
\right|-{1\over 2}{\rm Li}_3\left(e^{-2\left|a_{ij}\right|}\right)\right)+
{N_c\over 2}\sum_{i>j>k}a_ia_ja_k=
\\={1\over 4}\sum_{i,j}\left({1
\over 3}\left|a_{ij}^3
\right|-{1\over 2}{\rm Li}_3\left(e^{-2\left|a_{ij}\right|}\right)\right)+
{N_c\over 6}\sum_{i}a_i^3
\ee

The analogous expression with massive hypermultiplets included reads as
\cite{M}
\be\label{prep5}
F={1\over 4}\sum_{i,j}f^{(5)}(a_{ij})-{1\over 4}\sum_{i,\alpha}
f^{(5)}(a_i+m_{\alpha})+{1\over 16}\sum_{\alpha,\beta}
f^{(5)}(m_{\alpha}-m_{\beta})+\\+{1\over 12}(2N_c-N_f)
\left(\sum_{i}a_i^3+{1\over 4}\sum_{\alpha}m_{\alpha}^3\right)
\ee
where we have introduced the functions $f^{(d)}(x)$ equal to
\be\label{fd}
f^{(4)}(x)=x^2\log x,\ \ \ f^{(5)}(x)=\left({1
\over 3}\left|x^3
\right|-{1\over 2}{\rm Li}_3\left(e^{-2|x|}\right)\right)
\ee

Note that the second derivative of the $5d$ prepotential, which is the
effective coupling constant, is of the form $\log(\sinh aR_5)$. It can be
obtained from the second derivative of the $4d$ prepotential $\log(a)$ via
summing over the Kaluza-Klein modes $\log(\sinh aR_5)=\sum_m \log(a+m/R_5)$.

One can also calculate the perturbative prepotential for the $6d$ theory
via summing the effective charges over the Kaluza-Klein states in order
to get $f^{(6)}$:
\be
f^{(6)}(a)'' sim \sum_{nm}\log(a+\frac{n}{R_{5}}+\frac{m}{R_{6}})\sim
\log\theta\left(a\left|{R_5\over R_6}\right)\right.,\\
f^{(6)}(a)=\sum_{m,n}f^{(4)}(a+\frac{n}{R_{5}}+\frac{m}{R_{6}})=
sum_n f^{(5)}(a+n{R_{5}\over R_6})=\left({1
\over 3}\left|x^3\right|-{x^2\over 4}+{1\over 120}
-{1\over 2}{\rm Li}_{3,q}\left(e^{-2|x|}\right)\right)
\ee
where ${\rm Li}_{3,q}(x)$ is the elliptic tri-logarithm \cite{mamont,sk}.
This result should be compared with the manifestly calculated prepotential
\cite{M}.

Now let us discuss expression (\ref{prep5}) for the prepotential.
First of all, one can restore the dependence on $R_5$ by substitution $a_i\to
a_iR_5$ and $m_{\alpha}\to m_{\alpha}R_5$. Now we can looks at different
limits of our system.

The simplest limit corresponds to the $4d$ case and is described by
$R_5\to\infty$. In this limit, $f^{(5)}(x)
\stackreb{x\sim 0}{\to}f^{(4)}(x)$ and we reproduce the perturbative $4d$
prepotential. Moreover, comparing (\ref{prep5}) with the general expression
for the $4d$ prepotential (see \cite{wdvv})
\be
F = \frac{1}{4}\sum_{vector\ multiplets} \Tr_{A}
(\phi + M_nI_A)^2\log(\phi + M_nI_A) - \nn \\
- \frac{1}{4}\sum_{hypermultiplets} \Tr_R
(\phi + m_RI_R)^2\log(\phi + m_RI_R) + f(m)
\ee
($I_R$ denotes the unit matrix in representation $R$)
one can understand that the general result for the $5d$ prepotential reads as
\be
F = \frac{1}{4}\sum_{vector\ multiplets} \Tr_{A}
f^{(5)}(\phi + M_nI_A)
- \frac{1}{4}\sum_{hypermultiplets} \Tr_R
f^{(5)}(\phi + m_RI_R) + f(m)
\ee

Certainly, this $R_5\to 0$ limit can be easily reproduced at the level of
integrable system. Indeed, it suffices to make the replace $S_i\to R_5S_i$,
$\mu\to e^{R_5\mu}$ in the Lax operator (\ref{l-gen}) ((\ref{laxhg})) in
order to obtain the Lax operator of the $XXX$ (higher $SL(p)$) spin chain
(see \cite{ggm}).

The other interesting limit is the limit of the flat $5d$ space-time, i.e.
$R_5\to\infty$. In this limit, only cubic terms survive in the prepotential
(\ref{prep5}):
\be
F={1\over 12}\sum_{i,j}\left|a_{ij}\right|^3-{1\over 12}\sum_{i,\alpha}
\left|a_i+m_{\alpha}\right|^3+{1\over 48}\sum_{\alpha,\beta}
\left|m_{\alpha}-m_{\beta}\right|+{1\over 12}(2N_c-N_f)
\left(\sum_{i}a_i^3+{1\over 4}\sum_{\alpha}m_{\alpha}^3\right)
\ee
Therefore, now we discuss their meaning.

In fact, following N.Seiberg \cite{S}, we should identify these terms as
coming from the Chern-Simons (CS) Lagrangian. There are two different sources
of the cubic terms. The first one is due to the function $f^{(5)}(x)$.
Since this function can be obtained as the sum of the $4d$
perturbative contributions to the $f^{(4)}(x)$ over the Kaluza-Klein modes,
these cubic terms have perturbative origin and come from the 1-loop (due to
the famous effect of generation of the CS terms in odd-dimensional gauge
theories).

The second source of the cubic terms is due to the bare CS Lagrangian. As it
was shown in \cite{S,Sei2,IMS}, one can consider these terms with some
coefficient $c_{cl}$:
\be
{c_{cl}\over 6}\sum_i a^3_i
\ee
restricted only to satisfy the quantization condition $c_{cl}+{N_f\over
2}\in \Z$ and the inequality $\left|c_{cl}\right|\ge N_c-{N_f\over 2}$.

In formula (\ref{prep5}), this bound is saturated $c_{cl}=N_c-N_f/2$.
This value of $c_{cl}$ is actually distinguished, since only if
$c_{cl}=N_c-N_f/2$, the WDVV equations \cite{wdvv} are fulfiled. However, one
can obtain other values degenerating our integrable system. Indeed, let us
choose the special values of the second Casimirs at $m$ sites so that
$Q(\lambda)$ takes the form $\lambda^m\bar Q_{N_f}(\lambda)$, i.e. $m$
factors in (\ref{212}) become just $\lambda^m$. Then, one easily gets in this
case that $c_{cl}=N_c-m-N_f/2$. Taking into account that $m$ is integer and
the evident condition $m+N_f\ge 2N_c$, we immediately obtain the restriction
for $c_{cl}$ above. Note that the curve (\ref{sc2}) with this arbitrary $m$
coincides with that proposed in \cite{theisen}.

Note that now it is clear why the WDVV equations are not satisfied with $m\ne
0$. Since this corresponds to some degeneration of the integrable system,
it means that we fix some moduli (masses) w.r.t. which one should vary in the
WDVV equations (see \cite{wdvv} for details). Therefore, these latter are no
longer hold. Remark that it is assumed above that the prepotential is
defined in the fixed Weyl chamber. However, we leave the wall crossing jumps
out of the discussion.

\section{Solution of $5d$ theories via M theory}
Similar to $4d$ case, the $5d$ theories also have
different avatars, and, in the present section, we consider different
points of view on these theories, which come from strings.
Though brane construction of field theories in five
dimensions is not as much useful as in $d=4$, it still remains quite
illustrative. In what follows we derive the
general polynomial structure of the spectral curve in terms of the
toric diagrams. This data does not fix the dependence on moduli but fits
the information which can be obtained from the integrability. Moreover, it
turns out that, in five dimensions, the type IIB brane construction naturally
encodes the toric data of the M-theory compactification and, therefore, has
even more to do with its geometrical cousin. Note that the integrable
system actually encodes two ingredients -- the spectral curve and the linear
bundle on it. In what follows, we deal only with brane interpretation of the
spectral curve without any treatment of the linear bundles. The discussion on
the possible meaning of the linear bundles in brane terms can be found in
\cite{ggm,gor}. Remind briefly the geometrical construction of such theories
\cite{V1,eng,nikita3}.

\subsection{Geometrical engineering of $5d$ theories}

Five-dimensional SUSY gauge theories may be constructed via
M-theory compactification on Calabi-Yau threefold in the
vicinity of singular points \cite{IMS,nikita3,wmf}. Being interested
only in this region of parameters, we will discuss exclusively a local
model implying that its embedding into a compact Calabi-Yau is
always possible \cite{nikita3}.
To this end, we consider $A_{n-1}$ ALE spaces
fibered over a set of the base spheres, each sphere corresponding
to an $SU(n)$ gauge theory. Size of the base sphere
$V \propto {1 \over g^2}$ is governed by the coupling constant of
the corresponding $SU(n)$ factor. Kahler moduli of
$P^1$'s associated with blowing up the ALE space define the Cartan
scalars $a_i$ on the Coulomb branch moduli space.
This non-compact Calabi-Yau space is a hypersurface in
the holomorphic quotient and
can be represented as a set of Kahler cones, or, in other words,
by the toric polyhedron $\De$.

For example, one can start from $\CN=(2,2)$ linear sigma model
construction of Calabi-Yau \cite{22W,AGM} and define $U(1)$
charge vectors of matter fields:
\be
{\bf Q^a}=(q^a_1, \ldots , q^a_k)
\nn
\ee
From the $\sig$-model point of view these vectors define
gauge-invariant variables, and from the Calabi-Yau side they
act on homogeneous coordinates in projective space and
fulfil $r$ linear relations
\be
\sum_J Q^a_J {\bf \nu_J} =0
\label{rel}
\ee
between vectors ${\bf \nu_J} \in {\bf N} = {\bf Z}^d$ which define
coordinate patches in each Kahler cone. In order to have resolved
canonical Gorenstein singularities vertices ${\bf \nu_J}$ must
lie on the hyperplane $H$:$\langle {\bf \nu_J}, {\bf h}
\rangle =1$ in lattice ${\bf N}$. Therefore,
in the sequel we will assume ${\bf h}=\(1,0, \ldots, 0\)$, and
take ${\bf \nu_J} = \( 1, \* \)$. Moreover, for a Kahler cone $\sig$ to
be smooth at the origin, it must satisfy ${\rm Vol} \[ \sig \] = 1$
\footnote{This requirement comes out in relation to the brane
constructions, and will be discussed in the next subsection.}.
Vertices ${\bf \nu_J}$ in lattice ${\bf N}$ form a toric polyhedron
$\De = \{ {\bf \nu_J} \}$.
For the toric variety to be smooth, $\De$ must be convex.
Thus, having defined the original Calabi-Yau in terms of
holomorphic quotients, we now turn to the solution
of the model via mirror symmetry.

As it was explained in \cite{eng,V1}, Kahler moduli of the
original Calabi-Yau specify Coulomb moduli of the low-energy
effective theory. Under mirror transform they get mapped to the complex
structure of the mirror manifold, which is just Riemann surface
$E$ having the same moduli space. There is a profound reason in
this symmetry, because the Riemann surface is the spectral curve
of the associated integrable system. For the five-dimensional
case under consideration, this is a family of relativistic
(or otherwise, quantum) generalisations of $XXX$ spin chain
familiar to us from $d4$ analysis \cite{nikita3,ggm}. These
are inhomogeneous XXZ spin chains and their deformations.
For some simple
limits they correspond to Ruijsenaars model (pure gauge theory).

In terms of toric data a mirror manifold is very simple
and is also defined by toric variety on the dual lattice ${\bf M}$
\cite{Batyrev},
where each dual Kahler cone $\tilde \sig \subset {\bf M}$
is spanned by
vectors ${\bf \mu} \in \tilde \sig$, such that for any ${\bf \nu} \in \sig$
they satisfy $\sum_{\a} \mu^{\a} \nu^{\a} \ge 0$. In terms
of charge vectors this mirror manifold is:
\be
\prod_J y_J^{q^a_J} =1
\nonumber
\ee
For the models we consider here it is just a spectral
curve of underlying integrable system.

\subsection{Branes and solution of the model}

One could naively get the appropriate brane pattern via T-duality
along $x^4$ from the corresponding type IIA picture \cite{W}.
This drives us to type IIB string theory \cite{ah} with
typical brane picture like those in \cite{ah} \footnote{Our
notations differ from those of \cite{ah} in the following
way: $x^{3,4} \ra x^{8,9}$, $x^5 \ra x^3$ and $x^6 \ra x^7$.}:

\begin{equation} \ \ \ \left\{
\begin{array}{c|cccccccccc}
      & 0& 1& 2& 3& 4& 5& 6& 7& 8& 9\cr
NS5   & +& +& +& +& +& +& -& -& -& -\cr
D5    & +& +& +& +& +& -& +& -& -& -
\end{array}\right.
\label{tabl5}
\end{equation}
with D5-branes instead of D4. But under these changes holomorphic
solution to the Laplace equation $(s = x^6+ix^{10}) = \log (v=
x^4+ix^5)$ transforms into $x^6= {1 \over 2} | x^5 | =
{x^5 \over 2} {\rm sign} (x^5)$. The shape of the branes is
closely related to the perturbative couplings $\tau_{ij} =
\d_i \d_j \CF$ \cite{wmf,IMS},
as it takes place in four dimensions \cite{W}.
Moreover, to have IIB supersymmetry broken only up to ${1 \over 4}$,
when charge $n$ D5-brane meets charge $k$ NS5 they must merge into
$(k,n)$ bound state with the slope $\tan \phi = {k \over n}$ on
$(x^5, x^6)$ plane \cite{ah}.

Despite bending of the branes we can surround the system by
three-spheres $S^{(x^6,x^7,x^8,x^9) \ra \infty}$
and $S^{(x^5,x^7,x^8,x^9) \ra \infty}$, integration over
which yield us the brane charges. Since always we are
interested in Coulomb branch only, the first integral must give
$k$ units of NS charge and the last one should be equal to
$n$ units of RR charge. Moreover, following \cite{W,ah}, we
divide moduli of brane configuration into two groups: those whose
deformation change asymtotic branes, and those that do not.
The first group of parameters (e.g. coupling constants, bare quark
masses) define the five-dimensional theory, while the rest are
dynamical moduli. This is good point to note a correspondence
with geometrical engineering approach \cite{eng,V1}. In that
language each link in the toric polyhedron $\De$ is associated
with a 1-cycle on the corresponding Riemann surface $E$. Periods
along 1-cycles in the compact part of $E$ (i.e. internal links
in $\De$) define the dynamical moduli whereas bare parameters
arise from 1-cycles that wrap non-compact part of $E$ (i.e.
related to the links on the boundary of $\De$). One can go
further and note that the number of such external links in $\De$
corresponding to some low-energy effective field theory coincide
with the number of infinite fivebranes in the brane construction
of the same theory. Is it just an occasion?
The answer is no, and this is only beginning of the story.

It turns out that the brane picture and the toric polyhedron
corresponding to a five-dimensional field theory are intimately
related to each other. To establish other connections let us
visualize brane diagrams in other terms. Any such configuration
depicts kinematics of two-dimensional ($x^5$, $x^6$) free
particles. Charge conservation in the verteces where branes merge
together is nothing but momentum conservation law for free
particles. Integer-valued momentum $p^{(5,6)} =(k,n)$ of
particle is just $SL(2,Z)$ charge of the brane. The
vertical branes in terms of \cite{ah}
carry $(1,0)$ charge (momentum) vector,
and horizontal ones -- $(0,1)$. Remember that the number of vertical (NS5)
branes specify number of gauge groups whereas the number of
horizontal (D5) branes defines the rank of each $SU$ factor.
Namely, for $SU(n)^k$ gauge theory there are $k+1$
NS-branes and $n$ D-branes in each stack. And, similiarly,
the number of vertical links in toric polyhedron,
i.e. height of $\De$ equals $k+1$ and the number
of horizontal links, i.e. its horizontal size is $n$.
At generic point on moduli space, where $\De$ is nonsingular
and all canonical Gorenstein singularities are resolved (i.e.
$\sig$ ${\rm Vol} \[ \sig \] =1$ for each Kahler cone), vertices of
$\De$ lie on two-dimensional integer lattice. And in these
conventions we can endow each link in $\De$ with interger-valued
two-dimensional vector $(k_{\de},n_{\de})$.

We will exploit the fact that type IIB brane configurations
for five-dimensional field theories are in one-to-one correspondence
with toric polyhedrons of Calabi-Yau M-theory compactifications
to the same theories in $d=5$. The exact relation between
the two is provided by identification of $(k_{\de},n_{\de})$
charge (momentum) vector with a link vector in $\De$.

This powerful tool allows to solve any five-dimensional
$\prod SU(n)$ gauge theory with arbitrary matter. In the
meantime, charge conservation condition in IIB brane costruction
is somehow related to resolution of singular structure of
Calabi-Yau. This is an interesting subject for future
investigations. As it was argued in \cite{ah}, brane
configurations of five-dimensional theories naturally
incorporate group structure of the theory. This group
structure seems to be one of the key points in the solution
and is of very importance from the integrable system point
of view. It strongly justifies our result that brane picture
and toric data are almost the same thing, because the toric
polyhedron is also known \cite{V1} to represent quiver diagram.

The toric polyhedron $\De$ for the
case of pure gauge $SU(n)$ theory can be found in \cite{V1}.
Adding a massive quark,
we break one link on the boundary of $\De$ into
two pieces and add one more internal link. Therefore, the
number of internal and external (representing mass of the quark)
links increase by one. In the brane picture the story is exactly
the same. Including a fundamental matter in the theory, we attach
another semiinfinite brane. Note, that we can do it in four
different ways -- between the pairs of four semiinfinite branes
already presenting in the configuration. And the same
possibilities can be found in the breaking one of the four
external links of $\De$ into two parts. Thus a new semiinfinite
brane divides one of the internal branes into two segments, i.e
it adds one finite brane and one semiinfinite (by itself), in
perfect agreement with geometrical picture.

Let us stress here that the ambiguity we have just encountered
is a feature of all five-dimensional theories (and may be
even of all four- and six-dimensional theories). Always we have a choice
of the way to attach a "matter" brane or to fiber our Calabi-Yau
space. The simplest examples of this ambiguity already appeared
in the literature \cite{eng,V1,nikita3}, where one had a choice,
for example, between two Hirzebruch surfaces $F_0$ and $F_2$. The
models correspond to two different five-dimensional theories, but
flow to the same four-dimensional theory \cite{nikita3}. So, in
five dimensions field theories inherit more information of the
Calabi-Yau, and, in particular, become fibration-dependent.
It has been already explained in section 2 that, from the point of view of
integrable systems, all these ambiguitites corresponds to some specific ways
of degeneration of some larger system (reference system in terminology of
\cite{ggm}). This all-containing system always turns out to be the most
general integrable system allowed by symmetry requirements (twisted $XXZ$ in
$5d$ case). Certainly, this system contains enough number of parameters that
can be specifically fixed to reproduce various field theory counterparts.

For $k$ $SU(n_i)$ factors ($\prod_{i=1}^k SU(n_i)$
case) one has to take a fibration
of $n_i$ spheres over each of $k$ base spheres.
The total number of external links is $2(k+1)$
as the number of semiinfinite branes in the brane configuration.
The number of finite branes is
$\sum_{i=1}^k \(3 n_i -1\) -1$ which matches exactly the number of
internal links in $\De$\footnote{Here the asymptotic freedom of
the corresponding gauge factors, i.e. condition $2n_i \ge n_{i-1}
+ n_{i+1}$, is essential. It means that no semiinfinite branes
intersect each other and toric polyhedron is smooth (convex).}.

Now, having the directions for solving the model and strong
evidence for them, we review several solutions of
five-dimensional gauge theories.

\subsection{Some solutions}

Let us first brielfly remind the solution of pure gauge theory
in terms stated above. The $SU(2)$ low-energy gauge dynamics can be
described by brane configuration
(\ref{tabl5}) \cite{ah}. These configurations correspond to different choices
of fibrations, namely, to $F_0$ and $F_2$ Hirzebruch surfaces.
One of them provides the convensional
relativistic Toda system.

The toric variety is defined by vectors \cite{V1,nikita3}:
\be
{\bf \nu_J} =
\pmatrix {0 & 0 \cr -1 & 0 \cr 1 & 0 \cr  0 & -1 \cr   0 & 1}
\nonumber
\ea
that obey linear relations (\ref{rel}) with charge vector
\be
Q^a = \pmatrix { -2 & 1 & 1 & 0 & 0 \cr -2 &  0 & 0 & 1 & 1}
\nonumber
\ee
By this data one can easily derive the spectral curve for pure
$SU(2)$ gauge theory:
\be
a(t + {1 \over t}) + b (v + {1 \over v}) +1 =0
\nonumber
\ee
where $a$ and $b$ are parameters, related to the moduli and
radius of the space-time circle .
Vacuum expectation values of the fields (local coordinates
on the moduli space of the Coulomb branch) are periods of the
differential descending from the Calabi-Yau to this curve
in the (typical for all five-dimensional theories) form
\cite{nikita3}:
\be
\la= {1 \over 2 \pi i} log(t) {dv \over v}
\label{diff}
\ee
It was natural to expect $\la$ in the form (\ref{diff}) from the
very beginning where we described brane shape by the solution
of the Laplace equation in the appropriate dimension.

Generalization to the $SU(N)$ case is straightforward. The toric

polyhedron $\De$ corresponding to the brane configuration
can be found in \cite{V1}.
The ${\bf \nu}$ vectors for $SU(N)$ gauge theory are:
\be
{\bf \nu_J} = \pmatrix { -1 & 0 \cr 1 & 0 \cr  0 &
- \[ {n \over 2} \] \cr \ldots & \ldots \cr   0 & \[ {n \over 2} \]}
\nonumber
\ee
where square brackets denote smaller integer number.
The same phenomenon happens for the
theories with odd number of matter hypermultiplets
\cite{IMS,wmf}. The charge vector reads in this case:
\be
{\bf Q}=\pmatrix{q^1_J \cr q^2_J \cr \ldots \cr q^n_J} =
\pmatrix { 1 & 1 & -1 & 0 & 0 & 0 & \ldots & 0 & 0 & -1 \cr
0 & 0 & 1 & -2 & 1 & 0 & \ldots & 0 & 0 & 0 \cr
\ldots & \ldots & \ldots & \ldots & \ldots &
\ldots & \ldots & \ldots & \ldots & \ldots \cr
0 & 0 & 0 & 0 & 0 & 0 & \ldots & -2 & 1 & 0 \cr
0 & 0 & 0 & 0 & 0 & 0 & \ldots & 1 & -2 & 1}
\nonumber
\ee
It leads to the spectral curve:
\be
(t + {1 \over t}) + v^{- \[ {n \over 2} \]}P^{(n)}(v) =0
\label{sun}
\ee
where $P^{(n)}(v)$ is a $n$-degree polynomial in $v$.

We can include a matter hypermultiplet of mass $m$ in the
fundamental representation to our $SU(n)$ gauge theory by
attaching a semiinfinite brane to the right or to the left
of the brane configuration we have studied. Following arguments
of the previous subsection, in the geometrical setup we
add a non-compact divisor, so that:
\be
{\bf \nu_J} = \pmatrix { 1 & 1 \cr -1 & 0 \cr 1 & 0 \cr  0 &
- \[ {n \over 2} \] \cr \ldots & \ldots \cr   0 & \[ {n \over 2} \]}
\nonumber
\ee

Relations (\ref{rel}) remain unchanged, if we put:
\be
{\bf Q}=\pmatrix{q^1_J \cr q^2_J \cr \ldots \cr q^n_J} =
\pmatrix {
1 & 1 & 0 & 0 & -1 & 0 & 0 & \ldots & 0 & 0 & -1 \cr
0 & 1 & 1 & -1 & 0 & 0 & 0 & \ldots & 0 & 0 & -1 \cr
0 & 0 & 0 & 1 & -2 & 1 & 0 & \ldots & 0 & 0 & 0 \cr
\ldots & \ldots & \ldots & \ldots & \ldots & \ldots &
\ldots & \ldots & \ldots & \ldots & \ldots \cr
0 & 0 & 0 & 0 & 0 & 0 & 0 & \ldots & -2 & 1 & 0 \cr
0 & 0 & 0 & 0 & 0 & 0 & 0 & \ldots & 1 & -2 & 1}
\nonumber
\ee

Constructing the mirror manifold, as before, we find the
spectral curve:
\be
(v t + {(v - m) \over t}) +
v^{1 - \[ {n \over 2} \]}P^{(n)}(v) =0
\nonumber
\ee

Similiar arguments lead to the solution of theory with arbitrary
matter content. Thus, for $n_f$ quarks the corresponding
spectral curve is:
\be
(v^{n_f} t + {\prod_{l=1}^{n_f} (v - m_l) \over t}) +
v^{n_f - \[ {n \over 2} \]}P^{(n)}(v) =0
\label{sunf}
\ee

Note the $t \leftrightarrow {1 \over t}$ symmetry of
introducing "matter term" into (\ref{sunf}). It is a corollary of
the same symmetry in the way we can attach semiinfinite branes
or choose ALE fibration, as in four-dimensional case.

One can easily check that curve (\ref{sunf}) enjoys R-symmetry
and all global symmetries. This can be regarded as a good
consistency test for the spectral curves we obtain in this
letter.

Another consistency check is that all the theories we consider
here flow to their four-dimensional counterparts when $S^1$
factor of the space-time shrinks to zero size, $R \ra 0$.
In that limit all five-dimensional degrees of freedom decouple
and spectral curves (\ref{sun}), (\ref{sunf}) transform
into the corresponding Seiberg-Witten solutions.

We will not write down the full toric data for group product
case, as it can be directly, but tediously derived.
Brane configuration (\ref{tabl5}) for
the $\prod_{i=1}^k SU(n_i)$ gauge theory
is found in \cite{ah}, and corresponding
toric polyhedron -- in \cite{V1}.
The spectral curve for such theory is:
\be
t^{k+1} + \sum_{i=1}^k t^i v^{- \[ {n_i \over 2} \]}
P^{(n_i)}(v)+1=0
\nonumber
\ee
Note that one can introduce additional
parameters in the diagram.They are labeled by a set of integer
numbers $\de_i$, $i=1 \ldots k$. Each number $\de_i$ shifts the $i+1$-th row
of toric polyhedron by $\de_i$ units to the right if positive or to the left
in the other case. But deformations are not arbitrary, we still have to
preserve smooth structure of Calabi-Yau, i.e. $\De$ must be convex. This
implies a constraint $| \de_i | \le |n_i-n_{i-1} |$.  Taking deformations
into account, we end up with the family of spectral curves:  \be t^{k+1} +
\sum_{i=1}^k t^i v^{- \[ {n_i \over 2} \] + \de_i} P^{(n_i)}(v)+1=0 \nonumber
\ee

There are straightforward generalizations to other gauge groups
with arbitrary matter content. Taking any brane configuration for
a given gauge theory from \cite{ah}, one can associate
a toric variety with it and its
mirror.

\section{6d issues}
In this section we propose a general integrable framework of the $6d$ theories
with two compact dimensions. The main claim is that the $6d$ theory with
fundamental matter is naturally governed by the $XYZ$ spin chain on
$N_{c}$ sites. We get the explicit expressions for the corresponding spectral
curve (Riemann surface) and for the proper differential $dS$ that, in
paticular, determines the prepotential for the low-energy effective
action. An important feature of the $6d$ theories is that they
are self-consistant only if $2N_{c}=N_{f}$, otherwise, the theory becomes
anomalous \cite{6d1}-\cite{6d4}.

The $6d$ theory is related to the elliptic curve that comes from the
compact fifth and sixth dimensions. In what follows, we identify the modulus
of this spectral curve with the modulus of the elliptic curve that defines
the structure constants of the Sklyanin algebra. This latter is known to be
the symmetry group behind the $XYZ$ spin chain. In the previous $5d$
consideration we really dealt with the degenerated version of the Sklyanin
algebra, namely the quantum group with the parameter $q=\exp(-R_{5})$.

Remark that the most general integrable systems under would
correspond to the both coordinates and momenta defined on two different
tori \cite{duality}. In $5d$ theory without adjoint matter,
one deals with the degenerated situation and both phase
variables are defined on the cylinders, while in $6d$ gauge theory the
momenta of integrable system live on the torus with radii
$R_{5},R_{6}$, while the coordinates live on the cylinder.

Now we briefly describe the $XYZ$ chain that is proposed to be
in charge of the
particular $6d$
theories following the paper \cite{xyz}. The Lax operator of this chain
is defined on the elliptic curve
$E(\tau)$ and is explicitly given by (see \cite{FT} and references therein):
\be\label{39}
L^{Skl}(\xi) = S^0{\bf 1} + i\frac{g}{\omega}\sum_{a=1}^3 W_a(\xi)S^a\sigma_a
\ee
where
\be
W_a(\xi) = \sqrt{e_a - \wp\left({\xi}|\tau\right)} =
i\frac{\theta'_{11}(0)\theta_{a+1}\left({\xi}\right)}{\theta_{a+1}(0)
\theta_{11}\left({\xi}\right)}\\
\theta_2\equiv\theta_{01},\ \ \ \theta_3\equiv\theta_{00},
\ \ \ \theta_{4}\equiv\theta_{10}
\ee
Let us note that, for the sake of convenience and to keep the similarities
with formulas of \cite{xyz}, we redefine in this section the spectral
parameter $\xi\to i{\xi\over 2K}$, where $K\equiv\int_0^{{\pi\over
2}}{dt\over\sqrt{1-k^2\sin^2t}}={\pi\over 2}\theta_{00}^2(0)$,
$k^2\equiv{e_1-e_2\over e_1-e_3}$ so that $K\to{\pi\over 2}$ as $\tau\to
i\infty$. This factor results into additional multiplier $\pi$ in the
trigonometric functions in the limiting cases below.

The Lax operator (\ref{39})
satisfies the Poisson relation (\ref{quadr-r}) with the
numerical {\it elliptic} $r$-matrix $r(\xi)={i{g\over\omega}}\sum_{a=1}^3
W_a(\xi)\sigma_a\otimes\sigma_a$, which
implies that $S^0, S^a$ form the (classical)
Sklyanin algebra \cite{Skl1}:
\be\label{sklyal}
\left\{S^a, S^0\right\} = 2i\left(\frac{g}{\omega}\right)^2
\left(e_b - e_c\right)S^bS^c
\nn \\
\left\{S^a, S^b\right\} = 2iS^0S^c
\ee
with the obvious notation: $abc$ is the triple $123$ or its cyclic
permutations.

The coupling constant  ${g\over \omega}$ can be eliminated by simultaneous
rescaling of the $S$-variables and the symplectic form:
\be
S^a =\frac{\omega}{g}\hat S^a \ \ \ \
S^0 = \hat S^0\ \ \ \
\{\ , \ \} \rightarrow -2\frac{g}{\omega}\{\ ,\ \}
\ee
Then
\be
L(\xi) = \hat S^0 {\bf 1} +i\sum_{a=1}^3 W_a(\xi)\hat S^a\sigma_a
\ee
\be\label{sklyaln}
\left\{ \hat S^a,\ \hat S^0\right\} = -i
\left(e_c - e_b\right) \hat S^b\hat S^c \nn \\
\left\{ \hat S^a,\ \hat S^b\right\} = -i\hat S^0\hat S^c
\ee
One can distinguish three interesting
limits of the Sklyanin algebra: the rational, trigonometric and
double-scaling limits \cite{xyz}. We are interested in only
trigonometric limit here, since it describes the degeneration to the
$5d$ case. In this limit,
$\tau\rightarrow +i\infty$ and the Sklyanin algebra
(\ref{sklyaln}) transforms to
\be\label{trisklya}
\{ \hat S^3,\hat S^0\} = 0,\ \
\{\hat S^1,\hat S^0\} =i\hat S^2\hat S^3,\ \
\{\hat S^2,\hat S^0\} =-2i\hat S^3\hat S^1\\
\{\hat S^1,\hat S^2\} =-i\hat S^0\hat S^3,\ \
\{\hat S^1,\hat S^3\} = i\hat S^0\hat S^2,\ \
\{\hat S^2,\hat S^3\} = -i\hat S^0\hat S^1
\ee
The corresponding Lax matrix is
\be\label{laxxxz}
L_{XXZ} = \hat S^0{\bf 1}-{\f \sinh\pi\xi}\left(\hat S^1\sigma_1+
\hat S^2\sigma_2+\cosh\pi\xi\hat S^3\sigma_3\right)
\ee
and $r$-matrix
\be
r(\xi)={i\over\sinh\pi\xi}
\left(\sigma_1\otimes\sigma_1+\sigma_2\otimes\sigma_2+
\cosh\pi\xi\sigma_3\otimes\sigma_3\right)
\ee
Now one can note that the algbera (\ref{trisklya}) admits the identification
$\hat S_0=e^{S_0^{trig}}+e^{-S_0^{trig}}$,
$\hat S_3=e^{-S_0^{trig}}-e^{S_0^{trig}}$. With this identification and
up to normalization of the Lax operator (\ref{laxxxz}), we finally get the Lax
operator (\ref{l-gen}) of the $XXZ$ chain with the Poisson brackets
(\ref{pois}).

The determinant
$\det _{2\times 2} \hat L(\xi)$ is equal to
\be
\det _{2\times 2} L(\xi) = \hat S_0^2 + \sum_{a=1}^3 e_a\hat S_a^2
- \wp(\xi)\sum_{a=1}^3\hat S_a^2
= K - M^2\wp(\xi) =  K - M^2 x
\ee
where
\be\label{casi}
K = \hat S_0^2 + \sum_{a=1}^3 e_a(\tau)\hat S_a^2 \ \ \
\ \ \ \
M^2 =  \sum_{a=1}^3 \hat S_a^2
\ee
are the Casimir operators of the Sklyanin algebra (i.e. Poisson
commuting
with all the generators $\hat S^0$, $\hat S^1$, $\hat S^2$,
$\hat S^3$). The determinant of the monodromy matrix (\ref{Tmat})
is
\be\label{54}
Q(\xi) = \det _{2\times 2} T_{N_c}(\xi) =
\prod_{i=1}^{N_c} \det _{2\times 2} \hat L(\xi - \xi_i) =
\prod_{i=1}^{N_c} \left( K_i - M^2_i\wp(\xi - \xi_i)\right)
\ee
while the trace $P(\xi) = \frac{1}{2}\Tr T_{N_c}(\xi)$
generates mutually Poisson-commuting Hamiltonians.
For example, in the case of the {\it homogeneous} chain (all $\xi_i = 0$
in (\ref{54}))
$\Tr T_{N_c}(\xi)$ is a combination of the
polynomials:
\be
P(\xi) = Pol^{(1)}_{\left[\frac{N_c}{2}\right]}(x) +
y Pol^{(2)}_{\left[\frac{N_c-3}{2}\right]}(x),
\ee
where $\left[\frac{N_c}{2}\right]$ is integral part
of ${N_c\over 2}$, and the coefficients of $Pol^{(1)}$ and $Pol^{(2)}$
are Hamiltonians of the $XYZ$ model.
The spectral equation (\ref{specurv}) for the $XYZ$ model
is of the form:
\be\label{specXYZ}
w + \frac{Q(\xi)} w = P(\xi),
\ee
where for the {\it homogeneous} chain $P$ and $Q$
are polynomials in $x = \wp(\xi)$
and $y =   \frac{1}{2}\wp'(\xi)$.
Eq. (\ref{specXYZ}) describes the double covering of the elliptic
curve $E(\tau)$:
with generic point $\xi \in E(\tau)$ one associates the
two points of ${\cal C}^{XYZ}$, labeled by two roots $w_\pm$
of equation (\ref{specXYZ}).
The ramification points correspond to
$w_+ =w_- = \pm\sqrt{Q}$, or
$Y = \frac{1}{2}\left(w - \frac{Q}w\right) =
\sqrt{P^2 - 4Q}= 0$.

Note that, for the curve (\ref{specXYZ}),
$x = \infty$ is {\it not} a branch point,
therefore, the number of cuts on the both copies of $E(\tau)$ is $N_c$ and the
genus of the spectral curve is $N_c+1$.

Rewriting analytically ${\cal C}^{XYZ}$ as a system of equations
\be\label{cxyz}
y^2 = \prod_{a=1}^3 (x - e_a), \nn \\
Y^2 = P^2 - 4Q
\ee\label{hdb}
the set of holomorphic 1-differentials on ${\cal C}^{XYZ}$ can be chosen as
\be
v = \frac{dx}{y},\ \ \\
V_\alpha = \frac{x^\alpha dx}{yY} \ \ \
\alpha = 0,\ldots,
\left[\frac{N_c}{2}\right], \nn \\
\tilde V_\beta = \frac{x^\beta dx}{Y} \ \ \
\beta = 0, \ldots,
\left[\frac{N_c-3}{2}\right]
\ee
with the total number of holomorphic 1-differentials
$1 + \left(\left[\frac{N_c}{2}\right] + 1\right) +
\left(\left[\frac{N_c-3}{2}\right] + 1\right) = N_c+1$
being equal to the genus of ${\cal C}^{XYZ}$.

Given the spectral curve and the integrable system one can immediately
write down the ``generating" 1-differential $dS$.
It can be chosen in the following way
\be\label{dsxyz}
d\Sigma ^{XYZ} \cong d\xi \cdot \log w
\nn \\
dS^{XYZ} \cong \xi {dw\over w} = - d\Sigma ^{XYZ} + d(\xi\log w)
\ee
Now, under the variation of moduli (which are all contained in $P$,
while $Q$ is moduli independent),
\be
\delta(d\Sigma ^{XYZ}) \cong \frac{\delta w}{w}d\xi =
\frac{\delta P(\xi)}{\sqrt{P(\xi)^2-4Q(\xi)}} d\xi
= \frac{dx}{yY}\delta P
\ee
and, according to (\ref{hdb}), the r.h.s. is a {\it holomorphic}
1-differential on the spectral curve (\ref{specXYZ}).

To get the pure gauge $6d$, one would bring all masses of the
fundamental matter to infinity with imply large both Casimirs
of the Sklyanin algebra at each site of the lattice. The
desired system governing the pure gauge case could be called elliptic
Toda chain. In fact, this is unclear how to perform this procedure, since
masses seems to take their values only within the restricted
torus parallelogram.

Let us mention that
in six dimensions renormalizable theories can be represented as
special fixed points in string theory compactifications, e.g. what
are called $S(N)$, $T(N)$, etc. \cite{6d1,6d2,6d3,6d4}
All such theories at low energies
behave like $U(N)$ gauge theory, but differ in the regularisation
scheme. It is not obvious that all the possibilities have clear
brane interpretation, because of the uncertainty of the
six-brane world-volume effective theory. There is a construction
by Karch et.al. \cite{6d1}
which seems a plausible candidate for the description
of $\CN=1$ supersymmetric QCD with $N_f= 2 N_c$ flavors. For this
particular case we assume the simplest regularisation via compactification
to four dimensions on a space-like torus. The complex structure of the
latter specify the parameter in Sklyanin algebra. There is an argument
for such an identification. The brane picture for the theories looks like
six-brane segments terminating on fivebranes\footnote{The pattern is
suspended in the $x^5$ - $x^6$ plane with all the branes alinged
along the $x^6$-direction}.

\section{Conclusion}
In this paper, we complete the naive classification program for the
correspondences between the classical integrable systems with the
finite number degrees of freedom and the low-energy effective actions
for the SUSY YM theories in different dimensions. There is the perfect
agreement among possible deformations of integrable systems and
adding new parameters to field theory. In particular,
relativization and "double" relativization of the integrable system
is nothing but introducing into the game one or two additional
compact dimensions at the field theory side.

Actually, the reason for the sucsess of this identification is due to
the symmetries behind both theories. Any new parameter in integrable
system appears as a new "group like" parameter and we have seen above
that, along this way, we obtain quantum group and Sklyanin algebra. The
identification of these symmetries on the field theory side is very
promising but at a moment we can
approximately recognize them only at the level
of brane configurations encoding the relevant information about
the moduli space of the vacuum sector. In particular, in the $5d$ case
the Dynkin diagram structure in the brane language becomes especially
transparent. It would be interesting to recognize the hidden
symmetry of the vacuum sector in more conventional field theory
terms without reference to the toric diagrams for string
compactification.

Another question which was beyond the scope of this paper is
quantization of the integrable systems under consideration.
It is important that the quantization does not destroy the symmetry
structure of the problem. For instance, when quantizing
integrable system in the $R$-matrix formalism, one replaces
Poisson brackets by commutators without changing the group structure
constants. It is remarkable that, quantizating the integrable systems
relevant to the $6d$ theories, one would
deal with the generic parameter Sklyanin algebra without any
degenerations. In particular, there emerges the second parameter of
the Sklyanin algebra (Plank constant), which can not be seen at the classical
level and has the meaning of the position of the marked point on $T^{2}$.
This issue certainly appeals for some additional clarification.

The authors are grateful to S.Kharchev, A.Marshakov, A.Morozov,
A.Zabrodin.  The work of A.Gor. was partially supported by grants
RFFI-95-01-01101, CRDF-RP2-132, INTAS-93-0273, the work of S.Guk.  -- by Merit Fellowship in Natural Sciences and Mathematics and
grant RFBR-96-15-96939, the work of A.Mir.  -- by grants RFBR-96-02-16347(a)
and INTAS-93-1038.

\end{document}